\documentclass[runningheads]{llncs}

\usepackage[utf8]{inputenc}

\usepackage{hhline} \usepackage[usenames,dvipsnames]{xcolor}
\usepackage{diagbox}
\usepackage{breakurl}
\usepackage{eucal}

\usepackage{centernot}
\usepackage{mdframed}

\usepackage{paralist}
\usepackage{mathtools}
\usepackage{url}
\usepackage{extarrows}

\usepackage{etoolbox}
\usepackage{xspace} % Smart spaces after commands

\usepackage{stmaryrd} % for llbracket and rrbracket\textbf{\textbf{}}

\usepackage[inline,shortlabels]{enumitem} 
\newlist{inlinelist}{enumerate*}{1}
\setlist*[inlinelist,1]{%
  label=(\roman*),
}

\usepackage{xifthen}  % Extended conditional commands
\usepackage{ifthen}

\usepackage{nicefrac}

\usepackage{color}
\usepackage[usenames,dvipsnames]{xcolor}
\usepackage[most]{tcolorbox}

\usepackage{tikz} % for HSL
\usepackage{float}
\usetikzlibrary{shapes.geometric, arrows}

\usepackage{stackengine} % for xrightarrow with closer labels

\usepackage{pifont}% http://ctan.org/pkg/pifont

\usepackage{hyperref}
\usepackage{cleveref}

\usepackage[final,nomargin,inline,index]{fixme} % Simplified management of FIXME's
\fxusetheme{color}

\definecolor{Black}{HTML}{000000}
\definecolor{Gray}{HTML}{808080}
\definecolor{Magenta}{HTML}{FF00FF}
\definecolor{RubineRed}{HTML}{ED017D}
\definecolor{ForestGreen}{HTML}{028A0F}
\definecolor{MidnightBlue}{HTML}{006795}
\definecolor{Plum}{HTML}{92268F}

\FXRegisterAuthor{bart}{anbart}{\color{magenta} {\underline{bart}}}
\FXRegisterAuthor{enrico}{anenrico}{\color{blue} {\underline{enrico}}}

\hypersetup{
  breaklinks   = true,
  colorlinks   = true, %Colours links instead of ugly boxes
  urlcolor     = blue, %Colour for external hyperlinks
  linkcolor    = blue, %Colour of internal links
  citecolor    = red   %Colour of citations
}
\hypersetup{final} % Needed to generate hyperlinks even in draft mode (ERROR IN MAC)

\usepackage{orcidlink}

\usepackage{listings}
\definecolor{listingBG}{HTML}{FFFFCB}%
\definecolor{listingFrame}{HTML}{BBBB98}%
\definecolor{listingLineno}{rgb}{0.5,0.5,1.0}%
\definecolor{LightGrey}{rgb}{0.975,0.975,0.975}
\usepackage{wasysym} 
\usepackage{newunicodechar}
\newunicodechar{€}{\euro}

\lstdefinelanguage{txscript}{
  commentstyle=\color{Gray},
  morecomment=[l]{//},
  morecomment=[s]{/*}{*/},
  classoffset=0,
  escapechar=\$,
  morekeywords={if,then,else,contract,skip,require,fun,return,pays,sig,transfer},
  keywordstyle=\color{Plum}\bfseries,
  classoffset=1,
  morekeywords={},
  keywordstyle=\color{MidnightBlue}\bfseries,
}

\lstdefinelanguage{solidity}{
  , basicstyle=\ttfamily\linespread{1.15}\footnotesize\lst@ifdisplaystyle\scriptsize\fi
  , commentstyle=\color{Gray}
  , morecomment=[l]{//}
  , morecomment=[s]{/*}{*/}
  , escapechar=\$
  , classoffset=0,
  , keywordstyle=\color{NavyBlue}\bfseries
  , morekeywords={assert,require,if,then,else,for,break,call,delegatecall,transfer,send,case, catch,continue,do,while,emit, new, return, revert, selfdestruct, try, with, throw, switch, suicide}
  , classoffset=1
  , keywordstyle=\color{YellowGreen}\bfseries
  , morekeywords={external, implements, import, interface, internal, library, payable, pragma, private, protected, public, pure, returns, super, using, view}
  , classoffset=2
  , keywordstyle=\color{blue}
  , morekeywords={function, constructor, contract, constant, struct, address, bool, receive, fallback, byte, bytes, bytes1, bytes2, bytes3, bytes4, bytes5, bytes6, bytes7, bytes8, bytes9, bytes10, bytes11, bytes12, bytes13, bytes14, bytes15, bytes16, bytes17, bytes18, bytes19, bytes20, bytes21, bytes22, bytes23, bytes24, bytes25, bytes26, bytes27, bytes28, bytes29, bytes30, bytes31, bytes32, enum, int, int8, int16, int24, int32, int40, int48, int56, int64, int72, int80, int88, int96, int104, int112, int120, int128, int136, int144, int152, int160, int168, int176, int184, int192, int200, int208, int216, int224, int232, int240, int248, int256, mapping, string, uint, uint8, uint16, uint24, uint32, uint40, uint48, uint56, uint64, uint72, uint80, uint88, uint96, uint104, uint112, uint120, uint128, uint136, uint144, uint152, uint160, uint168, uint176, uint184, uint192, uint200, uint208, uint216, uint224, uint232, uint240, uint248, uint256, var, void, ether, finney, szabo, wei, days, hours, minutes, seconds, weeks, years}
  , classoffset=3
  , keywordstyle=\color{Plum}\bfseries
  , morekeywords={balance, block, blockhash, instanceof, coinbase, difficulty, gaslimit, number, timestamp, data, gas, sig, now, tx, gasprice, origin}
  , classoffset=4
  , keywordstyle=\color{red}\bfseries
  , morekeywords={abstract}
}
\usetikzlibrary{shapes.geometric, arrows, positioning, shapes}

\tikzstyle{start} = [rectangle, rounded corners, minimum width=3cm, minimum height=1cm,text centered, draw=black, fill=red!30]
\tikzstyle{stop} = [rectangle, rounded corners, minimum width=3cm, minimum height=1cm,text centered, draw=black, fill=blue!30]
\tikzstyle{process} = [rectangle, minimum width=3cm, minimum height=1cm, text centered, draw=black, fill=orange!30]
\tikzstyle{decision} = [align=center, diamond, minimum width=3cm, minimum height=1cm, text centered, draw=black, fill=green!30]
\tikzstyle{arrow} = [thick,->,>=stealth]
\newcommand{\ifempty}[3]{%
  \ifthenelse{\isempty{#1}}{#2}{#3}%
}

\newcommand{\ifdots}[3]{%
  \ifthenelse{\equal{#1}{...}}{#2}{#3}%
}

\newcommand{\hidden}[1]{}

\makeatletter
\newcommand*{\itemequation}[3][]{%
  \item
  \begingroup
    \refstepcounter{equation}%
    \ifx\\#1\\%
    \else  
      \label{#1}%
    \fi
    \sbox0{#2}%
    \sbox2{$\displaystyle#3\m@th$}%
    \sbox4{\@eqnnum}%
    \dimen@=.5\dimexpr\linewidth-\wd2\relax
    \ifcase
        \ifdim\wd0>\dimen@
          \z@
        \else
          \ifdim\wd4>\dimen@
            \z@
          \else 
            \@ne
          \fi 
        \fi
      \@latex@warning{Equation is too large}%
    \fi
    \noindent   
    \rlap{\copy0}%
    \rlap{\hbox to \linewidth{\hfill\copy2\hfill}}%
    \hbox to \linewidth{\hfill\copy4}%
    \hspace{0pt}% allow linebreak
  \endgroup
  \ignorespaces 
}
\makeatother

\newcommand{\Real}[1]{\mathrm{Real}}

\newcommand{\codefont}{\fontsize{9}{9}\selectfont}
\newcommand{\code}[1]{{\tt\codefont{#1}}}
\def\etc{etc.\@\xspace}

\newcommand{\eg}{e.g.\@\xspace}
\newcommand{\ie}{i.e.\@\xspace}
\newcommand{\wrt}{w.r.t.\@\xspace}

  {}

\DeclareMathAlphabet{\mathbfsf}{\encodingdefault}{\sfdefault}{bx}{n}

\newcommand{\waldistrarrow}[1]{\approx_{\$}}

\definecolor{LightGrey}{rgb}{0.95,0.95,0.95}
\definecolor{keyword}{HTML}{7F0055}

\newlength\replength
\newcommand\repfrac{.1}

\newcommand\rulewidth{.6pt}
\newcommand\tdashfill[1][\repfrac]{\cleaders\hbox to \replength{%
  \smash{\rule[\arraystretch\ht\strutbox]{\repfrac\replength}{\rulewidth}}}\hfill}

\newcommand\tdotfill[1][\repfrac]{\cleaders\hbox to \replength{%
  \smash{\raisebox{\arraystretch\dimexpr\ht\strutbox-.1ex\relax}{.}}}\hfill}

\newcommand{\contrAdvC}[2]{\mathcal{C}} % computational contract advertisement

\newcommand{\solcmc}{SolCMC\xspace}

\newcommand{\toolname}{Neuroforger\xspace}
\newcommand{\speclang}{GATE\xspace}

\newcommand{\falseOut}{\emph{False}\xspace}
\newcommand{\trueOut}{\emph{true?}\xspace}

\newcommand{\falseGround}{FALSE\xspace}
\newcommand{\trueGround}{TRUE\xspace}

\newcommand{\myshorttitle}{Certified violation witnesses for smart contracts verification via LLMs}
\newcommand{\mytitle}{\toolname: certified violation witnesses for smart contracts verification via LLMs}

\title{\mytitle}
\titlerunning{\myshorttitle}
\author{Massimo Bartoletti\orcidlink{0000-0003-3796-9774} \and Enrico Lipparini\orcidlink{0009-0009-0428-4403}}
\institute{University of Cagliari, Cagliari, Italy}
\begin{document}

\maketitle

\begin{abstract}
Recent large language models (LLMs) incorporate reasoning capabilities
that allow them to perform well in predicting whether a smart contract respects a certain property, suggesting a complementary approach to traditional formal-methods-based techniques for smart contract verification.
However, the application of LLMs in such context has two major issues: 
1) properties expressed in natural language are intrinsically ambiguous, and 
2) answers returned by LLMs have no guarantee of correctness.
In this paper, we address both issues simultaneously by: 
1) introducing a new formal specification language  that extends Solidity with abstract types, 
and 
2) designing a workflow that combines LLMs with type checking and concrete execution to generate and  validate violation witnesses (\ie, counterexamples).
% leverages LLMs as counterexample generators and pairs them with type checking and concrete execution.
% guaranteeing that ...
%2) designing a procedure that leverages LLMs as counterexample generators, pairing them with type checking and concrete execution.
%The key idea is to represent a specification as a Solidity test containing ``holes''; filling these holes concretizes the test into an executable counterexamples for the target property.
The key idea is to represent a specification as a Solidity test with (existentially quantified) variables of abstract type;
%filling these holes 
finding an instantiation of these variables to concrete values (of the correct type)
concretizes the test into an executable counterexample (PoC) for the target property.
% Our procedure guarantees that its outputs are actual counterexamples to the given property.
%
We implemented our procedure in the tool \toolname, 
 experimentally evaluating it on a smart-contract verification dataset drawn from literature, obtaining promising results that demonstrate its potential applicability in the wild.
\end{abstract}
\section{Introduction}
\label{sec:intro}
%\enriconote{stress che tool di verifica sia certora che solcmc non certificano }

Smart contracts are self-executing programs deployed on a blockchain that regulate asset transfer between users, forming decentralized financial ecosystems that currently manage assets worth hundreds of billions of dollars.
Due to the immutability of the code after deployment, the absence of intermediaries, and the high value of assets managed, smart contracts are an attractive target for attackers.
This is witnessed by a long history of attacks, estimated to have caused more than 6 billions of dollars of losses~\cite{Chaliasos24icse}.

Formal verification offers a principled approach to prevent attacks, 
allowing developers to assess whether their code respects suitable specifications before deployment.
In practice, however, rigorous verification of smart contracts remains extremely challenging. 
Many violations arise from unforeseen interactions with other contracts, where attackers exploit subtle business logic errors that fall outside the scope of tools targeting predefined vulnerability patterns~\cite{Sendner24sp,Chaliasos24icse}. 
Furthermore, compared to other kinds of software, smart contracts introduce new  attack vectors, \eg those based on transaction ordering dependencies~\cite{Eskandari19sok,Babel23clockwork}, exploits of the economic mechanisms~\cite{Qin21fc,Zhou23sp}, and quirks of the smart contracts language and execution environment~\cite{ABC17post}.
State-of-the-art verification tools present several limitations, both in terms of expressivity (interesting strategic properties are not expressible) 
and in terms of soundness, as due to the complex semantics of programming languages used for smart contracts
--- most notably Solidity --- 
% especially in the case of the most adopted language -- Solidity --, 
tools often have to abstract from the actual semantics, leading to unsoundness.
Furthermore, tools are rarely able to produce certificates that validate their answers, leaving the user with no choice but to trust their process.

%peculiarities wrt "normal" program verification (violations often come from unpredicted interactions with other contracts)

%(mention Solidity)
%- semantica sporca, aspetti di basso livello che rendono molto complesso l'uso di strumenti di verifica tradizionale

%- possibilità di avere avversari col potere di selezionare/ordinare transazioni

%- immediato riscontro dell'attaccante (ci guadagna direttamente soldi)

%issues in current verification tools (unsoundness, limited expressivity)

%Recently, Large Language Models (LLMs) have been proposed as a potential complement to traditional formal verification techniques~\cite{BLP26fc}.
% \enriconote{si attacca abbastanza male (inoltre, gli unici lavori su smart contracts sono gli ultimi 2)}
Recently, several works have explored how Large Language Models (LLMs) can assist program verification for tasks such as  invariants generation~\cite{PRB24ase,WCY24ase,KSC24fmcad,lemur}, automatic code repair~\cite{TCJ25ast}, and autoformalization of specifications~\cite{liu2025propertygpt,CM25arxiv}.
Notably, in the context of smart contracts, it has been investigated whether LLMs can act as \emph{alternatives} to classical formal verification techniques~\cite{BLP26fc}.
Empirical evidence suggests that modern reasoning-oriented LLMs %(such as, \eg, GPT-5) 
%can reliably \enriconote{meglio evitare reliably che potrebbe essere attaccabile}
are rather good at anticipating the validity of properties stated in natural language and at providing plausible explanations for their answers, without requiring a fixed or formal specification language.
Nevertheless, this approach suffers from two fundamental and tightly coupled limitations:
\begin{enumerate*}[1)]
\item the natural-language specifications fed as input to the LLMs are intrinsically ambiguous, and
\item there is no guarantee that the answers provided by the LLMs are correct.
\end{enumerate*}

%however, two issues: 1) inherent ambiguity of specifications expressed in natural language, 2) impossibility to \emph{certify} the answers returned by the LLM.
%
In this paper, we aim to tackle both issues. We will keep the focus on Solidity (the most used smart contract language), but the presented framework can easily be adapted to a variety of other languages.
Our contribution is the following.
%in this paper we tackle both issues (remark che sono strettamente collegati)

First, we introduce  \speclang, a formal specification language based on an extension of Solidity augmented with abstract types %\enriconote{domain-specific?} types
%behavioural \bartnote{??} types %high-level types
%(``\emph{set of contracts}'' and ``\emph{sequence of transactions}''), 
(representing contracts, transactions, and variables).
We then frame the verification problem %of determining the validity of a property
 as the problem of finding a concretization of a \speclang-sentence (\ie, an  assignment that maps all abstract entities to concrete ones), that makes all the assertions in the sentence -- if any\footnote{By definition, a specification that contains no assertions is satisfied by all valid concretizations.} -- evaluate to true.
 In other words, we see abstract entities as existentially quantified variables, 
 and look at a satisfiability problem.
 If such an assignment exists, we say that the specification is violated; otherwise, that it is valid. 
 In the following, we use the term \emph{specification} as a synonym for \speclang-sentence,
 and we call assignments \emph{counterexamples}, or, 
 following the terminology commonly used in cybersecurity, \emph{Proofs of Concept} (\emph{PoCs}).

% TODO provare a dire qualcosa sui tipi di proprietà esprimibili?

We define a procedure that takes as input a \speclang specification, % so defined constitutes the input
%%is then instrumental to design a 
%of a procedure %, given a formal specification, 
and returns either a certified counterexamples (which witnesses that the specification is violated with absolute certainty) or ``\emph{true?}'' (meaning that it was not able find a counterexample). %, with no guarantee that the formula is valid).
This procedure leverages LLMs as the main reasoning engine responsible for the production of counterexamples.
Exact methods are then used to either certify the LLM-produced counterexample, or to refute it. 
In the latter case, the LLM is then re-queried together with an explanation that motivates why the counterexample was not accepted. 
%providing the LLM with some feedback that explains the refutation, ).
%
In this setting,
the task of the LLM
only amounts to find
%consists in finding 
a correct assignment, % to the free variables in the formula, 
where
here 
\emph{correct} means two things: \begin{enumerate*}[i)]
    \item that the assignment is indeed a concretization (i.e., abstract entities are mapped to concrete terms of the corresponding type), and
    %that each variable is assigned to a value of the corresponding type
    \item that the assignment makes all the assertions in the sentence evaluate to true.
 % constraints.
\end{enumerate*}
The first condition can be checked via type checking, while the second condition can be checked through the concrete execution of the obtained PoC.
If any of these conditions are not met, the LLM is instructed to refine the assignment produced, taking into account the reasons for the failure (as in a standard CEGIS loop%
 \footnote{CEGIS stands for \emph{Counterexample-Guided Inductive Synthesis}. 
 Note that \emph{counterexample} in this acronym has a different meaning than \emph{counterexample} in the rest of the paper (which is used as a synonym for PoC).}%
).%
%\bartnote{il linguaggio craftato per il primo issue è strumentale ad addressare il secondo issue}
%For the second issue, we address how to certify that an LLM-produced counterexample indeed violates the property. This amounts to checking that the returned assignment a) respects the correct types, and b) satisfies all the constraints.
%we focus sulle proprietà false che sono quelle che mostrano le violazioni

%In order to work with Solidity without abstracting from its full semantic 

We implemented our procedure in the prototype tool \toolname, which uses {GPT-5}~\cite{gpt5cutoff} to produce PoCs and the Forge~\cite{forge} testing environment for concrete executions; in this preliminary version, the tool asks the user to manually validate the type checking. 

To the best of our knowledge, this is the first work 
in the field of smart contracts
that studies how to leverage LLMs to construct certified counter\-examples/PoCs.
\iffalse 
%enrico: commento, direi che è più un rischio che un beneficio metterlo qua
In other domains, we are only aware of approaches that 
leverage machine learning techniques to boost formal verification (e.g. by hinting invariants or selecting tactics),
 or that use formal methods to verify LLM-generated code (see also \Cref{sec:related}); 
however, we are not aware of any approach that use LLMs \emph{as} verification tools and certify the correctness of their answers. %, guaranteeing soundness for negative answers. 
\fi

%
% formally certify LLM-produced counterexamples in the context of smart contracts, 
%and it envision to open the way to a neuro-symbolic approach to smart contract verification. 

% Foundry testing module (Forge), in particular, % faithfully 

%citare che è work in progress?

%itemize contributions?

\paragraph{Contributions.} We summarise our main contributions as follows:
\begin{enumerate}

\item We propose a specification language for Solidity called \speclang, % We achieve this 
defined by extending  Solidity itself with abstract types that represent unknown contracts, transactions, and variables.

\item We define the verification problem as the problem of finding an assignment of the abstract type variables to concrete values,
%\bartnote{abstract types to concrete values?}
 such that all the assertions in the specification are satisfied.
%If such assignment exists, the specification is violated; otherwise, it is valid.

\item We propose a procedure that leverages LLMs to search for satisfying assignments,
and combines type checking with concrete execution to certify the correctness of the 
%inferred
%\enriconote{non capisco bene cosa voglia dire inferred}
 assignments.

\item We implement our procedure in the prototype \toolname and experimentally evaluate it on a dataset based on~\cite{BLP26fc}, showcasing the feasibility of the proposed approach.

\item We make the tool and the dataset pubicly available on a github repository.~\cite{repo}

\end{enumerate}

%Learning to Disprove: Formal Counterexample Generation with Large Language Models

%Counterexample Guided Program Repair Using Zero-Shot Learning and MaxSAT-based Fault Localization

\section{Background and running example}
\label{sec:background}

In this~\namecref{sec:background} we provide essential background on smart contracts in Solidity, on their formal verification, and on recent developments on integrating LLMs in the verification process. 

\subsection{Smart contracts and Solidity}

Abstractly, a blockchain can be seen as a state machine that creates crypto-assets and allows users to exchange them programmatically through \emph{smart contracts}.
The state of this machine records which assets are owned by which accounts, and each transaction advances the system to a new state by transferring assets between accounts or, in some cases, creating new assets or new smart contracts.
Solidity is the main programming language for smart contracts, and it is supported by leading blockchain platforms such as Ethereum, BNB Chain and Avalance~\cite{Rosetta25fgcs}.
In these blockchains, transactions are issued by users who control \emph{externally owned accounts} (EOAs).
A transaction may either directly transfer the native currency (\eg, ETH on Ethereum) to another EOAs, or invoke a function of a smart contract, possibly transferring currency along with the call.
Contract functions are written in a procedural style, with a few domain-specific constructs such as the transfer of currency, mechanisms to revert transactions upon given conditions, and controls over computational resources through the gas mechanism~\cite{Gasol}.
Similarly to object-oriented languages, each deployed contract instance has its own persistent storage, which contains the state variables and data structures defining the contract state.
In addition to this storage, every contract account also has an associated balance of the native currency, which is maintained by the blockchain execution environment and automatically updated whenever transfers to or from the contract occur.

We illustrate Solidity through a simple Bank contract in~\Cref{lst:bank}.
The contract stores in a key-value map the \code{credits} associated to each bank's client. 
%An account must invoke the function \code{deposit} with an amount of ETH. 
When an account (identified by \code{msg.sender}) invokes the function \code{deposit}, 
the effect is twofold: 
\begin{inlinelist}
\item transfer an amount of currency (identified by \code{msg.value}) from the sender's balance to the contract's;
\item record the sender's additional credit in the map.
\end{inlinelist}
The \code{withdraw} function first decreases the number of credits and then transfers the amount of currency from the contract to the user.
\begin{figure}[tb!]
\begin{lstlisting}[language=solidity,caption={Solidity code for the Bank contract.},label={lst:bank}]    
contract Bank {
  mapping (address user => uint credit) credits; // records credits in ETH
  
  // receives ETH from users and increases their credits      
  function deposit() public payable { // payable => can receive ETH
    credits[msg.sender] += msg.value;
  }
  
  // sends ETH to users and decreases their credits 
  function withdraw(uint amount) public {
    // reverts transaction if amount exceeds the sender's credits
    require(0<amount && amount<=credits[msg.sender]);   
    credits[msg.sender] -= amount;
    // invokes the receive function of msg.sender & transfers amount ETH  
    (bool success,) = msg.sender.call{value: amount}("");
    // reverts transaction if the previous call does not succeed
    require(success);
  }
}
\end{lstlisting}
\end{figure}

% \enriconote{problema: con Bank1, il comportamento è molto non-deterministico, ogni tanto - tanto il gpt5 di FC quanto gpt5 con Foundry - danno True e ogni tanto False (con una poc giusta in questo caso); alternativamente ci sono altre versioni leggermente più complicate con altri bug in cui ritornano sempre False (altri bug che non impattano su questa proprietà, ma che evidentemente per qualche motivo portano gpt5 a ipotizzare che la proprietà sia falsa) }

\subsection{Formal verification of Solidity contracts}
\label{sec:background:verification-tools}

Bugs in smart contract code are particularly critical compared to those in traditional software.
Once deployed, a smart contract's code is immutable, meaning that programming errors cannot be patched or fixed retroactively. 
Moreover, the permissionless nature of public blockchains allows any user to interact with deployed contracts, often without explicit authorization. As a consequence, vulnerabilities can be directly exploited by adversaries to steal, lock, or otherwise tamper with the digital assets managed by the contract.

Recent empirical studies show that logical errors in smart contract code account for the vast majority of losses caused by real-world attacks \cite{Chaliasos24icse}.
Such errors are inherently difficult to detect automatically, as existing vulnerability detection tools focus on specific classes of bugs (\eg, reentrancy or arithmetic overflows) rather than on arbitrary deviations from intended contract functionality.
As a consequence, formal verification has emerged as a central and long-standing goal in the smart contract security auditing industry, offering the promise of reasoning about contract correctness at the level of intended semantics.

This growing interest is reflected in the emergence of several formal verification tools for Solidity proposed in recent years, some of which have achieved industrial adoption.
SolCMC~\cite{Solcmc} and the Certora Prover~\cite{certora} are two prominent examples in this space, each offering a distinct specification language, relying on different  verification techniques, and providing different outputs to the users.

SolCMC~\cite{Solcmc}, which is integrated into the Solidity compiler, allows users to 
express properties as assertions embedded in Solidity code. 
These assertions are interpreted as state invariants that must hold in all reachable contract states.
SolCMC translates the instrumented contract into a set of logical constraints, 
which are then discharged by an SMT/CHC solver (Z3~\cite{DeB08} or Eldarica~\cite{Eldarica}).
% If so, it produces a trace witnessing the violation.

The Certora Prover~\cite{certora,certora-wip}, unlike SolCMC, decouples the specification of the properties from the contract code, providing a domain-specific language (CVL) for specs.
A CVL spec defines constraints on the execution of a contract, and assertions that must be true in all states satisfying the given constraints.  
Certora compiles the Solidity contract and the CVL spec into a logical formula, and relies on SMT solvers to determine if the spec is satisfied in all contract states.

Both tools address the state explosion problem by approximating the Solidity semantics.
Such approximations, however, may introduce false negatives (\ie, a property holds but the tool classifies it as violated), and --- more critically --- false positives (the property is violated, but the tool reports it as valid).
Both kinds of erroneous classifications occur in practice and have been observed even for relatively simple smart contracts~\cite{BFMPS24fmbc}.

A crucial problem is that assessing the reliability of a classification provided by these tools is quite challenging.
In particular, when these tools claim that a property is satisfied, they do not provide any certification: confidence in the result rests solely on their presumed correct implementation and sound approximation of the Solidity semantics
(but, as observed before, such approximations are not always sound, limiting the trustworthiness of such positive results).

When claiming that a property is violated, the two tools follow different approaches.
\solcmc attempts to construct a counterexample, in the form of a sequence of transactions 
leading the contract to a state where an asserted invariant is violated. 
However, in our hands-on experiments with SolCMC on a large dataset of verification tasks, %~\cite{FCobservant-github},
%\bartnote{qui citerei l'articolo piuttosto che il repo (anche se nel papero FC forse non facciamo mai questa osservazione, giusto?)} 
%enrico: sì la facciamo anche nel papero
we observed that such counterexamples are produced only in a small fraction of the cases in which SolCMC reports a violation.~\cite{BLP26fc}
Even when counterexamples are outputted, they often lack the level of detail required to independently validate them through concrete PoCs.

% moreover, it may return FNs in certain cases in which it abstracts from the Solidity semantics~\cite{solcmc-main}.

% ON SOLCMC
%https://github.com/leonardoalt/cav_2022_artifact/blob/v1.0.2-extended/solcmc_extended_version.pdf
%Counterexamples and inductive invariants. When a verification target is disproved,SolCMC provides a readable counterexample describing how to reach the bug.In addition to the line of code where the verification target is breached, thecounterexample states the trace of transactions and function calls leading to thefailure along with concrete values substituted for the arguments, and the valuesof the state variables at the point of failure. When necessary, the trace includesalso synthesized reentrant calls that trigger the failure.
%
% https://docs.soliditylang.org/en/latest/smtchecker.html
% A counterexample may be given, however in complex situations it may also not show a counterexample. This result may also be a false positive in certain cases, when the SMT encoding adds abstractions for Solidity code that is either hard or impossible to express.

The Certora Prover provides counterexamples in the form of contract states that are claimed to violate the specified property. 
Unlike SolCMC, it does not attempt to ensure that such states are reachable: rather, it is responsibility of the property designer to require in the spec suitable invariants on the contract state so to rule away violations happening in unreachable states. 
As a consequence, determining whether a produced counterexample is spurious (false negative) or not often requires a time-consuming manual analysis~\cite{certora-docs-verification-reports}.

\subsection{Running example}
\label{sec:background:running-example}

\newcommand{\propline}[2]{\smallskip\begin{tcolorbox}[frame empty]{\code{#1}: {``\emph{#2}''}}\end{tcolorbox}\smallskip}

To illustrate our approach, we will use as a running example the Bank contract in~\Cref{lst:bank}, on which we want to verify the following simple property:

\propline{\hypertarget{prop:P1}{withdraw-assets-credit-others}}{after any non-reverting call to function \code{withdraw()}, the balances of any user but the sender are preserved}

Although this property may seem trivially valid, given that the \code{withdraw} function only updates the credits of \code{msg.sender}, this reasoning overlooks a subtle aspect of Solidity's execution model: the semantics of the \code{call} statement.
Besides transferring an amount of currency from the caller to \code{msg.sender}, this statement also triggers the invocation to a function of the callee (\ie, \code{receive}, or \code{fallback} if the former is not present~\cite{solidity-receive}).
For EOAs and for contract accounts where these functions are empty, this call has no effect other than accepting the incoming currency.
If this were always the case, the property \code{withdraw-assets-credit-others} would hold. 
However, the sender may itself be a contract \code{A} whose \code{receive} or \code{fallback} calls another contract \code{B}, which in turn calls the \code{deposit} function of the Bank contract. 
In this scenario, the property \code{withdraw-assets-credit-others} is violated, since the call to \code{withdraw} indirectly affects the balance of an account different from the original sender \code{A} (namely, contract \code{B}).
%spiegazione perché dice True quando ground truth è False

The symbolic verification tools considered above give divergent results on this verification task: SolCMC returns ``unknown'' when used with Z3 and ``False'' with Eldarica (without a counterexample), while the Certora Prover incorrectly reports the property as ``True''~\cite{BLP26fc}.

\subsection{LLMs as verification oracles for Solidity}
\label{sec:FC26}

In recent work~\cite{BLP26fc} we investigated the effectiveness of LLMs as predictors of the truth of properties of Solidity contracts, using  a large dataset of verification tasks as a basis for a systematic empirical evaluation.
While GPT-4's performance was in line with expectations and relatively modest (overall accuracy $\sim$70\%), GPT-5 achieved substantially higher performance metrics (accuracy $\sim$92\%) and consistently provided high-quality explanations, both in terms of correctness, completeness and coherence with the required task.

In the prompt used for our experiments in~\cite{BLP26fc}, we provided as input to the LLM the verification task (Solidity contract and property in natural language), and asked the LLM to provide a verification report including the outcome (True, False, or Unknown), 
an explanation of the reasoning used to reach that outcome, and --- for the tasks with a False outcome --- a %concrete 
counterexample  showing a state and/or a sequence of transactions that violate the property.

When applied to our running example in~\Cref{sec:background:running-example}, GPT-5 provided the verification report in~\Cref{fig:gpt5:output}.

\begin{figure}[t!]
\caption{GPT-5 output of~\cite{BLP26fc} for the query \code{withdraw-assets-credit-others}.}
\label{fig:gpt5:output}
% \begin{footnotesize}
\input{res_gpt5_FC_Bankv1}
% \end{footnotesize}
\end{figure}

Although the counterexample produced by GPT-5 is intuitively correct and coherent with our informal explanation in~\Cref{sec:background:running-example}, it cannot be regarded as an actual certificate of property violation.
In particular:
\begin{inlinelist}
\item the contracts to be deployed by the attacker and the transactions leading to  violation are only partially specified;
\item there is no guarantee that the suggested transactions produce the predicted outcomes.
\end{inlinelist}

At this point, considerable human effort is still needed to refine the sketched counterexample produced by GPT-5 into a concrete PoC witnessing the property violation.
In the remainder of the paper 
%we show how LLMs can be leveraged to significantly reduce this effort.
we propose a different approach that still leverages LLMs but %within a framework that 
allows for fully automatically generated certified PoCs.

%\section{Method}
%\label{sec:method}

%\enriconote{era "Framework"}
%cappello che introduce le due sottosezioni

%descrizione ad alto livello (schemino?)

%\subsection{Foundry-based Specification Language}

%\section{Solidity-Extended Specification Language (SESL)}
\section{The GATE specification language}
\begin{figure}[t!]
\begin{lstlisting}[language=solidity,caption={Example of valid specification in \speclang.},label={lst:exSESL}]   
import "forge-std/Test.sol"; // import testing features
abstract contract c;  // to be concretized

contract Foo { // the contract under verification
  function sendEther(address to, uint256 amount) {     
    address(to).call{value: amount}(""); // sends `amount` of wei to `to`
  }
}
contract Test_Foo is Test {
  function test_send_ether() { // abstract PoC
    Foo foo = new Foo(); // creates a new contract instance foo of type Foo
    vm.deal(address(foo), 10); // gives 10 wei to contract foo 
    uint256 bal_before = address(foo).balance;
    abstract transaction[] txs;  // here I can craft some transactions
    uint256 bal_after = address(foo).balance; 
    abstract uint256 diff;  // here I can instantiate some integer
    assert(bal_after - bal_before == diff);
  }
}
\end{lstlisting}
\end{figure}

We  define the \speclang (Generative Abstract Types Expressions) specification language as
 an extension of Solidity with abstract types.
The abstract types of our interest can be split into three classes:
\begin{itemize}
	\item base types (\eg \code{uint256}, \code{bool}, \etc);
 \item the type \code{contract} (contracts in Solidity include state variables and functions, similarly to classes in object-oriented languages~\cite{solidity-contracts});
 %``\emph{set of contracts}", representing finite sets of contracts (in Solidity, each contract defines its own type~\cite{solcmc-contract-types}), and
 \item the type \code{transaction}, which represents the invocation of a contract function from a user.

%``\emph{sequence of transactions}", representing finite sequences of statements that consist in contract calls (equivalently, transactions) 
\end{itemize}

Among these, only \code{transaction} is not  a native type in Solidity.
Rather, it can be seen a tuple of base types: 
two \code{address}es `\code{from}' (the caller) and `\code{to}' (the callee),
one \code{uint256} `\code{value}' (the amount of ETH sent along with the transaction), 
one \code{byte4} `\code{selector}' (the function being called), 
and one \code{bytes memory} `\code{params}' (a dynamically-sized byte array representing function parameters).

In practice, for calls performed within a contract 
(\eg, the \code{call} in~\Cref{lst:bank}),
the address \code{from} will always be equal to the address of the contract itself.
% , since a contract is not allowed to make function calls on behalf of someone else. 
When crafting counterexamples, however, we want more flexibility, \eg we want to  consider calls between \emph{arbitrary} accounts 
(\eg, in our running example, a call from a user to Bank, and a call from the attacker contract to Bank).

To achieve this flexibility, we leverage the Forge environment (the testing module of the Foundry toolkit~\cite{Foundry2025}),
which provides powerful testing features.
Among them, we rely in particular on the following:
\begin{itemize}

\item \code{vm.prank(address)}, which causes the next contract call to be executed as it were sent from the given address;

\item \code{vm.deal(address, uint256)}, which sets the balance of the given address to an arbitrary non-negative integer value.

\end{itemize}
%\enriconote{dire che estendiamo le seq of tx con queste due (ma vietiamo altra roba)}
%
The concretization of an abstract \code{transaction} will be either:
\begin{inlinelist}

\item a \code{vm.deal}, representing an external transfer of currency to an account;

\item a \code{vm.prank} coupled with a \code{call}, representing the invocation of a contract function from a given address;

\item the deployment of a contract (via the keyword \code{new}).

\end{inlinelist}

\Cref{lst:exSESL} shows a toy example of a \speclang specification. 
First, we declare an \lstinline[language=solidity]{abstract contract c} and define a concrete \lstinline[language=solidity]{contract Foo}, with only one function \code{sendEther} that transfers some \code{amount} of ETH to an address \code{to}.
Then, we define a concrete \lstinline[language=solidity]{contract Test_Foo} (inheriting the type \code{Test} from the Forge library).
In this contract, we create a new instance \code{foo}, which we provide with 10 currency units through \code{vm.deal}.
Then, we declare an abstract array of transactions \code{txs}
and use two variables \code{bal\_before} and \code{bal\_after}
to record the balance of \code{foo} before and after performing \code{txs}.
Finally, we declare an \lstinline[language=solidity]{abstract uint256 diff} and assert it to be equal to the difference between \code{bal\_before} and \code{bal\_after}. 
  
 %Then we have an existentially quantified variable \code{diff} of type \code{uint256} (we syntactically represent existential quantification with \mbox{``\code{= ...}''}).
 %Finally, we have a contract \code{Foo} with a function \code{doSomething} that contains a declaration  of a variable \code{seq\_tx} of type \code{sequence\_of\_transactions}, that can represent any finite sequence of contract calls. 
 
\begin{figure}[t!]
\vbox{%
\begin{lstlisting}[language=solidity,caption={An executable concretization of the abstract PoC in \Cref{lst:exSESL}.},label={lst:exConcrSESL}
]

contract C1 { receive() external payable { } } // can just receive wei

contract Foo {
  function sendEther(address to, uint256 amount) {     
    address(to).call{value: amount}(""); // sends `amount` of wei to `to`
  }
}
contract Test_Foo {
  function test_send_ether() { // concrete PoC
    Foo foo = new Foo();
    vm.deal(address(foo), 10); // sets balance of `foo` to 10
    uint256 bal_before = address(foo).balance;
    C1 c = new C1(); // creates a new contract `c` of type `C1`
    foo.sendEther(address(c), 1);
    uint256 bal_after = address(foo).balance; 
    uint256 diff = 1;  // instantiate diff to 1
    assert(bal_after - bal_before == diff);
  }
}
\end{lstlisting}}
\end{figure} 
 
A \emph{concretization} of a \speclang specification is a compilable Solidity contract where the abstract variables have been replaced by values of the corresponding types. 
%, and instantiating existentially quantified variables.
\Cref{lst:exConcrSESL} shows a concretization of the spec in~\Cref{lst:exSESL}.
Note that: 
\begin{itemize}

\item \code{c} is concretized into a contract \code{C1} with a \code{receive} function that just accepts incoming funds~\cite{solidity-receive};

\item \code{diff} is concretized into the integer \code{1}; 

\item \code{txs} is concretized into a sequence of two calls: 
\begin{enumerate*}[1)]
\item the deployment of a new instance \code{c} of a contract of type \code{C1},
\item a call to the function \code{sendEther} that transfers 1 wei from \lstinline[language=Solidity]{Test_foo} (referred to be \code{this}) to \code{c}.
\end{enumerate*}
 	
\end{itemize}

\begin{figure}[t!]
  \vbox{
    \begin{lstlisting}[language=solidity,caption={Specification of \code{withdraw-assets-credit-others} in \speclang.},label={lst:codeSpec}] 

import "forge-std/Test.sol";
import "../Bank.sol";

abstract contracts[] cs;

contract BankTest is Test {         
  Bank immutable bank;
  
  constructor() {
    abstract address bank_deployer;
    vm.prank(bank_deployer);
    bank = new Bank();
  }
  
  function test_withdraw_assets_credit_others_violation() public {
    abstract transaction[] txs;
    abstract address user;
    uint256 user_creditsBefore = bank.credits(user);
        
    abstract uint256 amount;
    abstract address sender;
    vm.prank(sender);
    bank.withdraw(amount); // should not revert
    
    uint256 user_creditsAfter = bank.credits(user);
    
    assertNotEq(sender, user, "user equal to sender");    
    assertNotEq(user_creditsBefore, user_creditsAfter, "cred. not changed");
  }
}

\end{lstlisting}
}
\end{figure}
 
%\enriconote{notare che contracts di base ne serve solo 1 e cambia poco la posizione, invece su seq of tx conta la posizione e ce ne possono essere più d'una; possibilità di predicare sulle seqtx?}

We define the verification problem for a \speclang spec as the problem of finding a  concretization such that, when executed, all the assertions evaluate to true.
%In this case, 
If such concretization exists, we say that the spec is \emph{violated}, otherwise that it is \emph{valid}.
For example, the spec in~\Cref{lst:exSESL} is violated, 
since executing its concretization in~\Cref{lst:exConcrSESL} passes the final assertion.

\Cref{lst:codeSpec} displays the \speclang spec of the property 
% \code{withdraw-assets-credit-others} 
considered for our running example in~\Cref{sec:background:running-example}.
The spec features a variable \code{cs} of type \code{abstract contracts[]} outside of the test contract body, which represents an arbitrary array of contract definitions.
%In the  test contract,  we have abstract type variables for: the address that deploys the contract (\code{bank\_deployer}), an arbitrarily long sequence of transactions (\code{txs}) after the deployment (this is crucial to characterize the set of reachable states), and 
In the  test contract,  we crucially have an abstract variable of type \code{transaction[]}, which characterizes the set of reachable states after the deployment of the \code{Bank} contract.
Then, an address \code{sender} calls the \code{withdraw} function, sending \code{amount} of wei (note that, if the call reverts, then the test fails, \ie this is equivalente to asserting that the call does not revert). 
Finally, we assert that an address \code{user} is different from \code{sender}, and that its credit have changed. 
If there is an assignment of the abstract type variables to concrete values so that all the asserts pass, this demonstrates a violation of the property.

\section{Producing certified counterexamples}
\label{sec:prodcex}
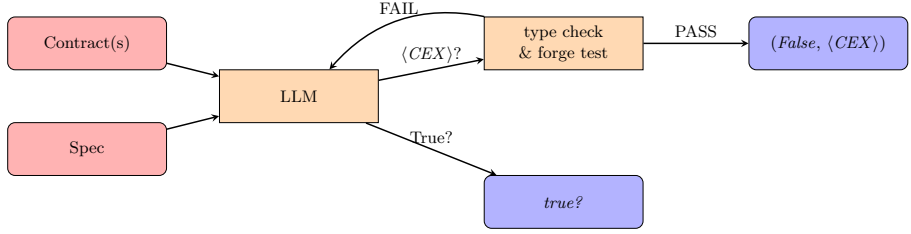
\begin{figure}[t!]
	\caption{Workflow of the procedure.}
	\label{fig:flowchart}
    \medskip
	\scalebox{0.7}{
		\begin{tikzpicture}[node distance=2cm, xshift=1cm, yshift=-1cm]
			% Nodes
			\node (contract) [start] {Contract(s)};
			\node (spec) [start, below of=contract] {Spec};
			\node (llm) [process, right of=spec, xshift=2cm, yshift=1cm] {LLM};
			\node  (typeforgecheck) [process, right of=llm, xshift=3cm, yshift=1cm] {\shortstack{type check \\ \& forge test}};
			\node (true) [stop, right of=llm, yshift=-2cm, xshift=3cm] {\trueOut};		
			\node (false) [stop, right of=typeforgecheck,  xshift=3cm] {(\falseOut, $\langle \mathit{CEX} \rangle$)};
			%\node (false) [process, left of=decision, xshift=-4cm] {FALSE};
			%\node (pass) [startstop, below of=true, yshift=-1cm] {PASS};
			%\node (fail) [startstop, below of=false, yshift=-1cm] {FAIL};
			
			% Arrows
			\draw [arrow] (contract) -- (llm);
			\draw [arrow] (spec) -- (llm);
			\draw [arrow] (llm) -- node[anchor=south]{$\langle \mathit{CEX} \rangle$?} (typeforgecheck);
			\draw [arrow] (llm) -- node[anchor=south]{True?} (true);
			\draw [arrow] (typeforgecheck) -- node[anchor=south] {PASS} (false);
			\draw [arrow] (typeforgecheck) edge[bend right] node[anchor=south] {FAIL} (llm);
			%\draw [arrow] (true) -- (pass);
			%\draw [arrow] (decision) -- node[anchor=south] {False} (false);
			%\draw [arrow] (false) -- (fail);			
		\end{tikzpicture}
	 }
 \end{figure}

We present in this~\namecref{sec:prodcex} our method to produce certified counterexamples.
The high-level workflow of our procedure is displayed in~\Cref{fig:flowchart}.

\begin{figure}[t!]
 \begin{lstlisting}[language=solidity,caption={A certified counterexample for the spec in \Cref{lst:codeSpec}, generated by GPT5.},label={lst:codeSolut}] 
  [...]  // imports
  contract User {
   function doDeposit(Bank bank) external payable {
    bank.deposit{value: msg.value}();
   }
   receive() external payable {}
  }
  contract Attacker {
   Bank public bank;
   User public user;
   
   constructor(Bank _bank) {
    bank = _bank;
   }
   
   function setUser(User _user) external {
    user = _user;
   }
   
   // When receiving ETH from Bank.withdraw, forward it to User,
   // which deposits into Bank, crediting the User.
   receive() external payable {
    user.doDeposit{value: msg.value}(bank);
   }
  }
  
  contract BankTest is Test {   
   constructor() {
    address bank_deployer = address(0xBEEF);
    [...]  // deployment of bank
   }
   
   function test_withdraw_assets_credit_others_violation() public {
    User u = new User();
    Attacker a = new Attacker(bank);
    a.setUser(u);
    vm.deal(address(a), 1 ether);
    vm.prank(address(a));
    bank.deposit{value: 1 ether}();
    
    address user = address(u);
    [...] // def. of user_creditsBefore
    uint256 amount = 1 ether;
    address sender = address(a);
    [...] // call to bank.withdraw and asserts
   }
  }
 \end{lstlisting}
\end{figure}
The procedure takes as input a set of contracts and a \speclang spec.
Then, it invokes an LLM, prompting it to produce a concretization of the spec, or --- if it believes no concretization exists --- to output ``\trueOut''
(the question mark signals that there is no guarantee that the answer is correct).
%
%The prompt adopts a zero-shot chain-of-thoughts approach; see Appendix~\ref{app:promptv2}.
%\bartnote{questa frase sul prompt meglio spostarla in experimental setup?}
%
% If the LLM is not able to produce a counterexample, the procedure returns \code{true?} 
%
When the LLM produces a tentative concretization, we: 
\begin{inlinelist}
    
\item check whether it is really so
(\ie, if its concrete types match the abstract ones);

\item execute the concretization: if it passes all the assertions, then it is an actual counterexample to the property, and we return it, together with ``\falseOut''.

\end{inlinelist}
If any of the two checks fail, the LLM is re-prompted,
receiving as input the original problem, its previous answer, and the reason why the answer was rejected.
% In this case, the user has the complete guarantee that the specification is indeed violated,  with a Proof-of-Concept that witnesses a violation.

The type checking of the concretization is used to prevent the LLM from cheating, \ie not respecting the abstraction restrictions.
For example, if the prompt fed to the LLM is not sufficiently precise, the LLM could \eg concretize an abstract integer variable to a concrete integer variable followed by a contract call, thus violating types. 

The procedure can be parameterised by $N_{\it iter} > 0$, which defines the maximum number of iterations between the LLM and the checks: if such limit is surpassed, the procedure returns ``\trueOut'', where the question mark is to remind that this prediction is not certified (unlike for the ``\falseOut'' case, where the answer is certified to be correct).

% Secondly, if the type check succeeds, we assess whether all the assertions in the generated counterexample pass, using the Forge testing module to perform a concrete execution of the code. 
% If any of the assertions fail, then the counterexample is not accepted. 

To exemplify of our procedure, in \Cref{lst:codeSolut} we show a certified counterexample to the specification in~\Cref{lst:codeSpec} generated by GPT-5. 
Note that LLM's answers are not deterministic, hence running again the procedure (even using the same LLM) may produce different outputs.
\section{Experiments}

%We have performed an initial experimental evaluation of our approach, gathering useful information about its feasibility.

\paragraph{Dataset.}

We base our experiments on the dataset in~\cite{BLP26fc}, 
focussing on the Bank use case described in~\Cref{sec:background:running-example}.
The dataset contains 17 versions of the Bank contract and 27 properties, which are common to all the versions.
Besides the original contract in~\Cref{lst:bank}, the other 16 versions are manually constructed by mutating the original code so to introduce logic errors or divergent behaviours that may affect the validity of some of the properties.

The properties included in the dataset are of different kinds~\cite{BCL25fmbc}: 
\begin{itemize}
\item \emph{function specs} (\eg, a non-reverting call to a certain function produces a certain effect on the state of the caller and the callee), 
\item \emph{state invariants} (\eg, the sum of all users' credits does not exceed the contract balance), 
\item \emph{metamorphic properties} (\eg, two different sequences of transactions produce the same
effect on the contract state),
\item \emph{strategic properties} (\eg, there exists a
sequence of transactions that produces some desired effect on the contract
state --- such as no ETH balance remains frozen in the contract).
\end{itemize}

For each verification task --- \ie, pair (property, version) --- the dataset includes a
\emph{ground truth}, \ie a boolean telling whether the property holds or not for that version.
In order to feature a $\sim$ 1:1 ratio between positive and negative tasks, a sub-dataset of 121 verification tasks is sampled.

\paragraph{Experimental setup.}

We implement the tool \toolname, following the  workflow depicted in \Cref{fig:flowchart}. 
We used GPT-5 as LLM model (although, in principle, the tool is modular \wrt the model).
The LLM prompt (which we do not include here for space constraints, but whose content is available on the public repository) adopts a zero-shot chain-of-thoughts approach.
We use the OpenAI API's default parameters at the time of the experiments: the sampling temperature is set to 1.0, nucleus sampling was disabled (top-p = 1.0), and no frequency or presence penalties were applied. 
The maximum number of generated tokens is 30,000, with generation terminating either upon reaching this limit or according to the model's internal end-of-sequence criteria. 
We set $N_{\it iter} = 3$ as the maximum number of iterations between the LLM and the checks.
For the type checking, we rely on manual validation, interactively asking the user to confirm whether the check passed or to provide an explanation for why it failed.
% We set to $3$ the maximum number of iterations between the LLM and the checks: if such limit is surpassed, the procedure returns ``\trueOut'', where the question mark is to remind that this prediction is not certified (unlike for the ``\falseOut'' case, where the answer is certified to be correct).
% For the type checking, we rely on manual validation, interactively asking the user to confirm whether the check passed or to provide an explanation for why it failed. 
%
The total cost of the experiments was below €10.

%We consider all the 27 properties available for that use case, formalizing them, whenever possible, in \speclang.  

%\enriconote{qualcosa su disambiguazione properietà a fix ground truth quando necessario?}

\paragraph{Expressibility.}

We encode in \speclang 22 of the 27 properties for the Bank use case featured in the dataset.
In particular, we observe that \emph{state invariants}, \emph{single-transition invariants}, and \emph{multiple-transition invariants} are, in general, naturally expressible in \speclang.
Also the \emph{metamorphic} properties present in the dataset (\ie, properties that predicate over different paths, in the sense of temporal logic~\cite{Chen18csur}) are expressible in \speclang, using the Forge feature of ``snapshots''~\cite{forge-snapshot}.

Of the 22 encoded properties, 2 required more advanced specifications using ``diff recording''~\cite{forge-diff-rec}, a feature of Forge that allows the inspection of the content of transactions.
 %\begin{enumerate*}[1)]
Indeed, in order to specify in \speclang, \eg, that ``\emph{the balance of the contract is greater than or equal to the sum of all users' credits}'', one needs to track the sum at run-time. 
This is only possible by writing a function that, given a sequence of transactions, checks whether the users credits have been changed, and, if so, keeps track of the changed values (this is analogous to hooks in CVL~\cite{certora-hooks}).
A similar trick is needed to predicate over internal function calls (\ie  calls triggered by other calls, rather than fired by an EOA).
In this case, the specification must inspect the content of a transaction at the low-level to assess whether the given function  has been called internally or not. 
This can also be achieved in \speclang through diff recording.
%\end{enumerate*}

As a side note, it would have been possible to express in \speclang properties about call interleavings (\eg, ``\emph{if \code{f1} is called after \code{f2}, no matter how many other calls have been made in between, then $P$ holds}''). 
However, no property of this kind is included in the dataset~\cite{BLP26fc} for the Bank use case. 
We note that this kind of properties is usually out of the scope of formal verification tools, as it requires to predicate over arbitrarily long sequences of transactions, whereas  standard specification languages are usually limited to a fixed number of transactions.
%\bartnote{stona ``traditional'' con ``transactions''}
In principle, it might still be possible to prove some of these properties, by writing (and proving) inductive invariants that constraint the changes that the blockchain state can incur after a transaction (hence generalizing to changes after an arbitrarily long sequence of transactions).
In practice, however, discovering such invariants can be extremely challenging and it can require lots of manual work.
 %\enriconote{check}

Of the 5 properties for which we do not provide \speclang specifications, 3 seem expressible in \speclang, but only at the cost of writing very complex  specifications.
These properties have the form ``\emph{there exists a unique address such that $P$ holds}'', for different predicates $P$.
Writing a counterexample to such properties would either require to show two distinct addresses such that $P$ holds (which is easy to express in \speclang), or show that there is no address such that $P$ (\ie, for all addresses $\neg P$).
The latter would be essentially inexpressible, given that the number of possible addresses is $2^{256}$, which makes it unfeasible to loop over all addresses.
For certain properties, including those in our dataset (where, \eg, $P$ has the form ``\emph{the address balance has changed}''), one could restrict to the addresses that occur in a sequence of transactions, relying on the fact that $P$ trivially holds for all the other addresses.
Therefore, a \speclang specification could be written by looping over these addresses.
This approach, however, would require the user to ensure that all the relevant addresses have been properly tracked; a subtle flaw may cause the specification to diverge from the property intended meaning.
For this reason, we opted for not including such specifications in our experiments.

% (intuitively, one would have to keep track of all the addresses that are mentioned in a sequence of transaction, and then loop over them, ).

The remaining 2 properties for which we do not provide a \speclang specification represent so-called liquidity/enabledness~\cite{Solvent,Schiffl24fmbc}, having the form ``\emph{in every state, it is \emph{always} possible to perform some actions to reach a state in which $P$ holds}''. 
Properties of this kind are not expressible in \speclang, in general.
Indeed, here the issue is that, in order to show a violation, one needs to \emph{prove} that, \emph{for all possible actions}, it is not possible to reach a state in which $P$ holds. 
This negative answer is more similar to a proof, rather than to a counterexample/PoC.
We note that, apart from academic tools that only work on a restricted fragment of Solidity~\cite{Solvent,KindHML},
liquidity is out of the scope of all other provers.

%
% TODO mettere motivo altre 3: 
%1 forall address
%2 loop over maps
%per la liquidity potrebbe essere comunque interessante studiare l'uso di llm per produrre witness (ovvero strategie) che riducono la liquidity a multiple transition invariant, che poi si può dimostrare con un tool di verifica

\begin{table}[t]
	\begin{center}
		\begin{tabular}{|c|c|c|}
			\hline
			\backslashbox{Output}{Ground truth}& \ \ \trueGround \ \ & \ \ \falseGround  \ \ \\
			\hline
			\trueOut & 57 & 4\\
			\hline
			\falseOut & 0 & 48\\
			\hline
		\end{tabular}
	\end{center}
	\caption{Experimental results.}
	\label{tab:results}
\end{table}

\paragraph{Results.}
Our experiments encompass 110 verification tasks, of which 57 with positive ground truth and 53 with negative ground truth.
Results are displayed in~\Cref{tab:results}.
Crucially, we see that the tool never returned \falseOut for verification tasks that have a  positive ground truth. 
Among the verification tasks with negative ground truth, the tool returned a counterexample for 48 instances, 
while for the remaining 5 instances it did not.
For 4 of these 5 instances, we validated the ground truth by manually constructing a valid concretization of the \speclang specification.
The remaining instance is quite interesting: a ``valid'' counterexample exists, in theory, but it requires $2^{256}-1$ function calls, making it unfeasible in practice.
Given the ambiguity on the ground truth for this verification task,
we believe that the output of the tool should not be considered incorrect in this case, hence we discard this verification task when computing metrics. % (as we have shown it to be for the other 4 instances).

We furthermore report in \Cref{tab:results_iter}  the data on the number of iterations performed by the tool before returning an answer. 
All iterations are due to failures during concrete test executions; no types violations occurred in our experiments.
As we can observe, the tool never used all the 3 iterations allowed by the iteration limit, and it rarely used 2 iterations ($3\%$ of the verification tasks). 
For $14\%$ of the verification tasks, it performed 1 iteration, while for the remaining $85\%$ no iteration was performed. 
Iterations distribute almost proportionally between verification tasks with true and false ground truth.
This shows that the feedback returned by Forge can help the LLM both at understanding that a tentative violation cannot actually occur, and at refining a violation that is indeed possible although not yet precise.

\begin{table}[b!]
  \centering
  \begin{tabular}{|c|c||c|c|c|}
    \hline
    \backslashbox{\# iter.}{results}
    & { \quad Total \quad} 
    & {\  (\trueGround, \trueOut)\  } 
    & { \ (\falseGround, \trueOut) \ } 
    & {\ (\falseGround,\falseOut)\ } \\
    \hline
    Total & 109 & 57 & 4  & 48 \\
    \hhline{|=|=|=|=|=|}
    0 & 90 & 46 & 4  & 40 \\
    \hline
    1 & 15 & 10 & 0  & 5  \\
    \hline
    2 & 4  & 1  & 0  & 3  \\ % the zero corresponds to the verification task discarded due to ambiguity on ground truth
    \hline
    3 & 0  & 0  & 0  & 0  \\
    \hline
  \end{tabular}
  \medskip
  \caption{Experimental results, where each prediction is associated to the number of iteration where it is obtained. 
  The rows represents the three possible results based on the pairs \emph{(ground truth, output)}. 
  }
  \label{tab:results_iter}
\end{table}

\paragraph{Discussion.}

Overall, our technique shows very good performance metrics: an accuracy of $96\%$, a precision of $100\%$ (note that this is ensured by type checking and concrete execution), a recall of $93\%$,
% a specificity of $92\%$,  % commento perché è asimmetrico (dovremmo includere anche sensitivity) e disturba per il commento sotto
and an F1 score of $96\%$.
These metrics are in line with those obtained in~\cite{BLP26fc} on the same dataset by querying GPT-5 with a natural language property, and asking it to return True/False, together with a natural language explanation (see \Cref{sec:FC26}).
%
% A small quantitative difference can be appreciated, 
A closer quantitative comparison with~\cite{BLP26fc} reveals that our technique achieves higher precision and lower specificity. 
% bart: qui non possiamo veramente argomentare con i numeri, perche' i dataset sono diversi
This is easily explained by the fact that our procedure is guaranteed to return \falseOut only when the counterexample is certified, hence false negatives do not exist by construction.
Indeed, a key \emph{qualitative} difference, is that, thanks to the fact that the properties that we pass in input are unambiguous (being written in a well-defined specification language),
the counterexample returned by our procedure are guaranteed to actually violate the property.
Hence not only false negatives are avoided, but also true negatives are \emph{guaranteed} to be as such. 
While this might not seem too important when working in a controlled environment where the ground truth is known, 
it is of fundamental relevance \emph{in the wild}, where an incomplete or inconsistent answer could be of little value.
% (the certification of positive results, \ie proofs, is of a much higher degree of complexity in practice, but in principle it would be possible to achieve).
%

%\enriconote{qualcosa su numero di iterazioni? forse non necessario (idealmente si potrebbero voler inferire delle cose, ma visto il dataset ristretto andrei cauto)}

%\paragraph{Discussion.}

%\enriconote{...?}

% \enriconote{problemi che ci possiamo aspettare
% type checker (per ora manual)
% (nota che l'output non è deterministico)
% }

\section{Related work}
\label{sec:related}

%\label{sec:related}
%\enriconote{spostato da su} 
In the field of smart contracts, some works~\cite{Gervais25ai,Prompt2Pwn,PoCo} have experimented  with LLMs to generate PoC of exploits.
Compared to our work, ~\cite{Gervais25ai,Prompt2Pwn} only take in consideration a single fixed property, \ie the ability for an attacker to extract crypto-currency from a contract, and only with respect to snapshots of on-chain data.
%
%This limits the applicability of using LLMs assistance during the development phase.
On the contrary, we consider arbitrary properties expressible in a rich formal specification language, and arbitrary reachable blockchain states. 
While ~\cite{Gervais25ai,Prompt2Pwn} show how attackers can leverage LLMs to exploit on-chain contracts, %(\cite{Gervais25ai}, in particular, warns about the risk that the reasoning capabilities of LLMs could favor exploitation over defense).
we envision our approach as a defense technique,
providing an analysis that can be performed before deployment to assess whether desirable properties can be violated.
Similarly to our, PoCo~\cite{PoCo} takes the side of the defender, but it requires an auditor to manually write natural-language vulnerability descriptions that are then converted to executable PoCs by an LLM. 
Our approach, on the contrary, only asks  the user  to write the specification, and then relies on the LLM to discover the vulnerability.
Moreover, the PoCs produced by PoCo are not guaranteed to actually reflect the vulnerability descriptions passed as input, whereas, in our case, using a formal specification language prevents this from happening.
%\bartnote{si può aggiungere che PoCo non certifica la vulnerabilità, visto che la concretizzazione può divergere dalla specifica in linguaggio naturale (da noi questo non può succedere grazie al typing)}

%One general purpose approach is neural model checking~\cite{GKPT25neurips}, in which neural networks are 

Other general-purpose approaches leverage machine learning techniques to boost formal verification (\eg, by hinting invariants~\cite{PRB24ase,WCY24ase,KSC24fmcad,lemur}, selecting tactics~\cite{MachSMT,ijcai2025p887}, or instantiating quantifiers~\cite{BJO19aitp,JJPU26ijar}).
Other lines of research are concerned with the verification of LLM-generated code (\eg ~\cite{CM25arxiv}),
and with pairing LLMs with formal methods for automatic code repair~\cite{TCJ25ast}.

We are not aware of approaches that specifically aim at using LLMs in program verification to produce counterexamples. 
We conjecture this might be due to the fact that, in the smart contract domain, finding a counterexample usually requires to also generate code (of the attacker's contracts).
This might be a peculiarity of the field, in which, in order to successfully use LLMs for verification, they are required to, \emph{at the same time}, generate code and try to determine the validity of a property.	

% \bartnote{\cite{JinC23tdsc}} \enriconote{non mi sembra usi LLM (comunque forse interessante)}

%Very recently (18 December) Certificates in AI: Learn but Verify \url{https://dl.acm.org/doi/full/10.1145/3737447}

%tries to uses LLMs \emph{as} verification tools and certify the correctness of their answers. %, guaranteeing soundness for negative answers. 

%\ 

%https://arxiv.org/abs/2410.23790 Neural Model Checking
%\ 

%Lemur: Integrating Large Language Models in Automated Program Verification \url{https://arxiv.org/pdf/2310.04870}

%Fact Verification in Knowledge Graphs Using LLMs 
%\url{https://www.dei.unipd.it/~silvello/papers/2025-SIGIR_Demo_LLM.pdf} check

%\enriconote{direi sacrificabile ora come ora}

%Survey: \cite{GarfattaKGG21acsw}
\section{Conclusions}
\label{sec:conclusions}

We proposed a technique that leverages LLMs to generate counterexamples for verification tasks on Solidity smart contracts. 
The key feature of our technique is that it guarantees that, whenever a counterexample is produced, it certificates the violation of the property.
To this purpose, we introduced a specification language based on an extended version of Solidity with abstract types (representing, \eg, contracts and transactions), 
and framed the verification problem as the problem of finding a valid concretization that makes the assertions contained in the specification evaluate to true.
The task of the LLM consists then in producing such a concretization,
whereas its validity is certified via type checking,
and the satisfaction of the assertions via concrete execution.
To help the LLM find a correct concretization, when one of the checks fails, we recur to a CEGIS loop that provides the LLM with a useful feedback. 
We implemented the proposed workflow in the prototype tool \toolname, showcasing its effectiveness on a simple yet paradigmatic use case.

Our neuro-symbolic approach can be seen as a complement to traditional formal verification techniques, which usually struggle to produce counterexamples (sometimes leading to unsound results).

%We do not suggest our method to \emph{substitute} traditional formal verification tools (as it cannot guarantee the validity of a property), but rather act as a \emph{complement} able to ground property violations with executable Proof-of-Concepts. 

\paragraph{Future works.}

%We plan to advance and strengthen this work as follows:
There are several directions in which this work can be extended, such as:
%\begin{itemize}
	systematically evaluate \toolname over the full benchmark in~\cite{BLP26fc},
	implement a type checker to automatically certify the correctness of the concretization,
	and
	investigate  alternative prompting techniques (\eg, feeding to the LLM also a description of the property in natural language).
%\end{itemize}

Moreover, while the current paper only tackles the problem of producing certified negative answers (i.e. counterexamples), 
we believe that similar neuro-symbolic techniques %it will be possible 
can also be designed to tackle  the dual problem of producing certified positive answers (i.e. proofs).
A more long-term research goal is to understand what kind of workflow would be necessary in order to achieve automatic proof generation. 
%, starting from the specification language 

%applicare tool in modo sistematico su benchmark

%provare a vedere se passare proprietà in linguaggio naturale aiuta (1 / 2 step, prima chiedi cex a chiacchiere poi di produrre poc)

%rendere più usabile dsl per specifica proprietà

%type checker automatico

% studiare se possibile usare LLM per generare 

% usare LLM per produrre witness di prop di liquidity 

%While the current paper only tackles the problem of producing certified negative answers (i.e. counterexamples), 
%we believe that similar neuro-symbolic techniques %it will be possible 
%can also be designed to tackle  the dual problem of producing certified positive answers (i.e. proofs).

\paragraph{Acknowledgments}

Work partially supported by project SERICS (PE00000014)
under the MUR National Recovery and Resilience Plan funded by the
European Union -- NextGenerationEU, and by PRIN 2022 PNRR project DeLiCE (F53D23009130001).

\bibliographystyle{splncs04}
\bibliography{main}

%\appendix
%\input{promptv2.tex}

\end{document}